\documentclass[aps,prc,twocolumn,floatfix,showpacs]{revtex4}
\usepackage{amssymb}
\usepackage{amsfonts}
\usepackage{graphicx}
\begin{document}
\newcommand{\be}{\begin{equation}}
\newcommand{\ee}{\end{equation}}
\newcommand{\bea}{\begin{eqnarray}}
\newcommand{\eea}{\end{eqnarray}}
\newcommand{\f}{\frac}  
\newcommand{\la}{\lambda}
\newcommand{\ve}{\varepsilon}
\newcommand{\ep}{\epsilon}
\newcommand{\da}{\downarrow}
\newcommand{\up}{\uparrow}
\newcommand{\V}{{\cal V}}
\newcommand{\ovl}{\overline}
\newcommand{\Ga}{\Gamma}
\newcommand{\ga}{\gamma}
\newcommand{\bra}{\langle}
\newcommand{\ket}{\rangle}
\newcommand{\eff}{_{\rm{eff}}}
\newcommand{\av}{{\rm{av}}}
\newcommand{\fl}{{\rm{fl}}}
\newcommand{\ina}{{\rm{in}}}
\newcommand{\wt}{\widetilde}
\newcommand{\ov}{\overline}
\newcommand{\G}{{\cal G}}
\newcommand{\Ha}{{\cal H}}
\newcommand{\sig}{\sigma}
\renewcommand{\le}{\leqslant}
\newcommand{\De}{\Delta}
\title{Addendum: Attenuation of the intensity within a superdeformed 
band}
\author{A.~J.~Sargeant}
\author{M.~S.~Hussein}
\author{M.~P.~Pato}
\affiliation{Instituto de F\'\i sica, Universidade de S\~{a}o Paulo,
Caixa Postal 66318, 05315-970 S\~{a}o Paulo, SP, Brazil}
\author{N. Takigawa}
\affiliation{Department of Physics, Tohoku University, 
Sendai, 980-8578, Japan}
\author{M.~Ueda}
\affiliation{Akita National College of Technology, 
Iijima Bunkyo-cho 1-1, Akita, 011-8511, Japan} 
\date{January 28, 2004}
\begin{abstract}
We investigate a random matrix model 
[Phys. Rev. C {\bf 65} 024302 (2002] 
for the decay-out of a superdeformed band as a function of
the parameters: $\Ga^\da/\Ga_S$, $\Ga_N/D$, $\Ga_S/D$ and $\De/D$.
Here $\Ga^\da$ is the spreading width for the 
mixing of an SD state $|0\ket$ with a normally deformed (ND) doorway state
$|d\ket$, $\Ga_S$ and $\Ga_N$ are the electromagnetic widths of the the SD
and ND states respectively, $D$ is the mean level spacing of the compound
ND states and $\De$ is the energy difference between $|0\ket$ and $|d\ket$.
The maximum possible effect of an order-chaos transition is 
inferred from analytical and numerical calculations of the decay 
intensity in the limiting cases for which the ND states obey Poisson 
and GOE statistics. 
Our results show that the sharp attenuation of 
the decay intensity cannot be explained solely by an order-chaos transition.
\end{abstract}
\maketitle

In superdeformed (SD) bands the total intraband decay intensity of the
super-collective $E$2 $\gamma$ transitions disappears 
suddenly due to tunneling through the barrier separating the 
superdeformed (SD) and normally deformed (ND) minima.
\cite{Wilson:2003,Lauritsen:2002,Paul:2001,Dewald:2001}. 
The theoretical calculation of the spin at which the decay-out occurs 
for different mass regions and the steepness of the attenuation of the decay
intensity are subject to uncertainties concerning the density
of ND states and the parameters describing the 
deformation barrier and collective motion \cite{Yoshida:2001}.
In Ref.~\cite{Aberg:1999a} \AA berg suggested an alternative explanation 
of the sharp decay-out: an order-chaos transition in the ND states 
enhances the tunneling probability and consequently the decay-out
is a manifestation of ``chaos assisted tunneling''.

In Ref.~\cite{Sargeant:2001aq} the authors investigated \AA berg's suggestion
by calculating the decay intensity as a function of the 
chaoticity parameter which produces a transition from order to chaos.  
We found that increasing the chaoticity did not enhance 
the decay out and concluded
on this basis that the decay-out must be due to the spin dependence of the 
barrier.
Subsequently, \AA berg \cite{Aberg:2003} criticised our assumption of an
energy difference of zero between the decaying SD state and the ND doorway 
state to which it is assumed to decay. In the following we study the
decay intensity as a function of the energy difference and as function
of the other parameters relevant to the decay-out, calculating the 
decay intensity in the limits that the ND states obey Poisson and GOE
statistics.
This permits us to infer the maximum possible effect that an order-chaos 
transition in the ND states can have on the decay intensity.
Our results reinforce our belief that the decay-out is mostly due to the
spin dependence of the barrier.

\begin{figure}
\includegraphics[width=.48\textwidth]{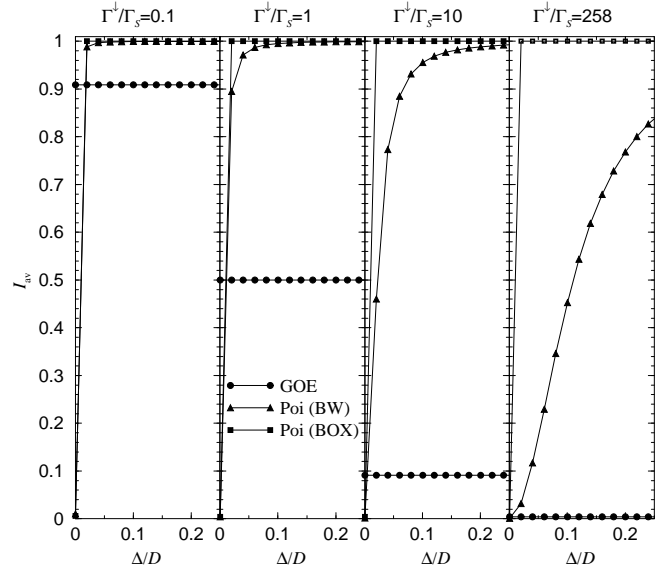}
\caption{\label{fig1} Decay intensity $I_\av$ vs. $\De/D$. 
The circles were calculated using 
Eq.~(\ref{IavGOE}), the triangles by substituting Eq.~(\ref{BW}) 
into Eq.~(\ref{1Iav}) and the squares using Eq.~(\ref{IavBOX}).
We set $\Ga_N/D=3\times 10^{-4}$ and $\Ga_S/D=6\times 10^{-6}$ which
are the relevant values for $^{194}$Hg-1 at
spin 12 $\hbar$ \cite{Krucken:2001we}.}
\end{figure}
The total average intra-band decay intensity of an SD band is given by 
\cite{Sargeant:2001aq}\footnote{Eq.~(\ref{Iav}) is in fact the background 
contribution to the average decay intensity. 
The fluctuation contribution should be added to $I_\av$
\cite{Gu:1999bv,Sargeant:2002sv,Hussein:2003ia}. 
Further, it appears possible that the 
variance of the decay intensity contains useful information about the decay-out
mechanism \cite{Gu:1999bv,Sargeant:2002sv,Hussein:2003ia}.}
\bea\nonumber
&&I_\av=\f{\Ga_S}{2\pi}\int_{-\infty}^{\infty}dE
\\&&\f{1}{[E-E_0-2\pi|V_{0d}|^2R_d]^2
+[\Ga_S+2\pi|V_{0d}|^2S_d]^2/4}.\hspace{4mm}
\label{Iav}
\eea
The intermediate SD state in the two-step 
decay which Eq.~(\ref{Iav}) describes is denoted $|0\ket$ and has energy $E_0$.
The electromagnetic width for the intra-band decay is $\Ga_S$. 
In what follows we assume that
$|0\ket$ only mixes (by tunnelling through the barrier in deformation 
space separating the SD and ND wells) with one special ND doorway
state $|d\ket$ whose energy is $E_d$.  The interaction energy
of $|0\ket$ and $|d\ket$ is $V_{0d}$. 
The state $|d\ket$ is subsequently mixed by the residual interaction 
with the remaining ND states, $|Q\ket$, $Q=1,...N$, 
having the same spin as $|0\ket$ and $|d\ket$. 
This strong doorway assumption was called {\em model B} in 
\cite{Sargeant:2001aq}. 
The $|Q\ket$ lie in the interval 
$L=ND$ where $D$ denotes the mean spacing in energy of the $|Q\ket$.
The functions $R_d(E)$ and $S_d(E)$ describe the manner in which $|d\ket$ is 
distributed in energy over the remaining ND states and are given by
\be\label{Rd}
R_d(E)=\f{1}{2\pi}\sum_{Q=0}^N
|c_d(Q)|^2\f{E-E_Q}{(E-E_Q)^2+\Ga_N^2/4}
\ee  
and 
\be\label{Sd}
S_d(E)=\f{1}{2\pi}\sum_{Q=0}^N
|c_d(Q)|^2\f{\Ga_N}{(E-E_Q)^2+\Ga_N^2/4}
\ee   
respectively, where $\Ga_N$ is the electromagnetic width the ND states.

\begin{figure}
\includegraphics[width=.48\textwidth]{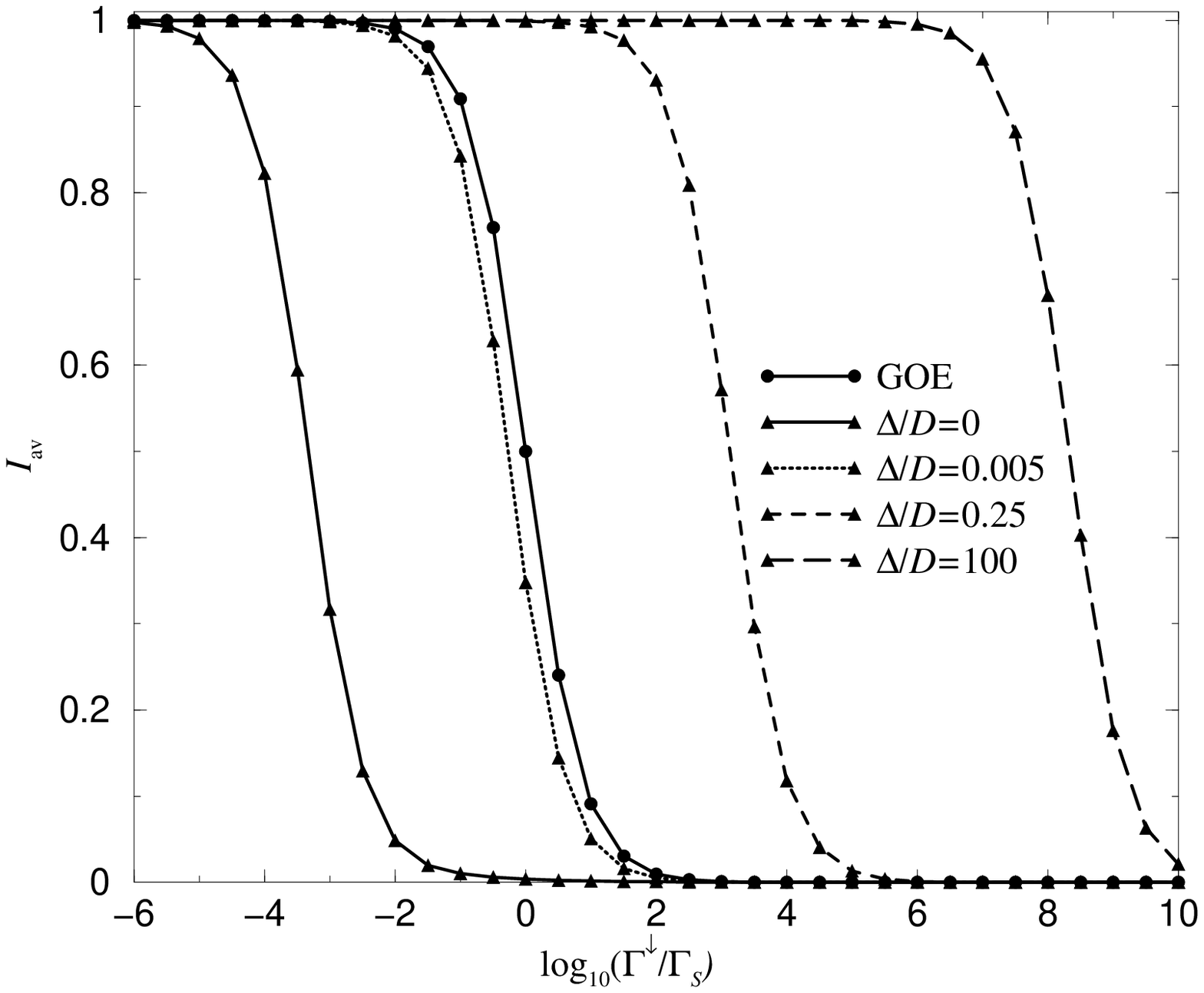}
\caption{\label{fig2} Decay intensity $I_\av$ vs. 
$\log_{10}(\Ga^\da/\Ga_S)$.
See also caption to Fig.~\ref{fig1}.}
\end{figure}
In \cite{Sargeant:2001aq} the effect of the chaoticity 
of the ND states on $I_\av$ was investigated by varying the strength 
of the residual interaction of the $|Q\ket$ and their interaction with 
$|d\ket$, both being assumed to be proportional to a parameter $\la$ 
(the chaoticity parameter) 
which may be varied continuously in the range $0<\la\le 1$. 
The limiting value $\la=0$ results in the $|Q\ket$ having
Poisson statistics (regularity) while $\la=1$ results 
in their having GOE statistics (chaos).
The value of $\la$ determines the shape of $S_d(E)$ [and $R_d(E)$]
which is precisely the strength function that was investigated as a 
function of $\la$ in \cite{Sargeant:1999qk}. 
In Ref.~\cite{Aberg:2003} it was pointed out that the calculations
of \cite{Sargeant:2001aq} were restricted to $E_d = E_0$.
We now study the Poisson limit of {\em model B} of
\cite{Sargeant:2001aq} for $E_d\ne E_0$.

As $\la\rightarrow 0$ and $\Ga_N\rightarrow 0$, 
$S_d(E)\rightarrow\delta(E-E_d)$. For non-zero $\la$, $S_d(E)$ broadens
with increasing $\la$ until when $\la=1$ it takes a form well 
approximated by
\cite{Sargeant:1999qk}
\be\label{SGOE}
S_d^{\rm{GOE}}(E)=\left\{
\begin{array}{cc}
1  ,&|E-E_d|\le L/2\\
0  ,&|E-E_d| >  L/2
\end{array}
\right..
\ee
Inserting Eq.~(\ref{SGOE}) into Eq.~(\ref{Iav}) we find 
that \cite{Weidenmuller:1998xf} 
\be\label{IavGOE}
I_\av^{\rm{GOE}}=(1+\Ga^\da/\Ga_S)^{-1},
\ee
as long as $\Ga_S+\Ga^\da\ll L$.

Instead of studying the interpolation between the limits $\la=0$ and
$\la=1$ by numerically diagonalising random matrices and performing
ensemble averages as was done in \cite{Sargeant:1999qk,Sargeant:2001aq},
we restrict ourselves to the limiting case $\la=0$ and 
use two representations of $\delta(E-E_d)$ broadened by $\Ga_N$:
the Breit-Wigner function,
\be\label{BW}
S_d^{\rm{BW}}(E)=\f{1}{2\pi}\f{\Ga_N}
{(E-E_d)^2+\Ga_N^2/4},
\ee
and the box function,
\be\label{BOX}
S_d^{\rm{BOX}}(E)=\left\{
\begin{array}{cc}
2/(\pi\Ga_N),&|E-E_d|\le \pi\Ga_N/2\\
0  ,&|E-E_d| >  \pi\Ga_N/2
\end{array}
\right..
\ee
\begin{figure}
\includegraphics[width=.48\textwidth]{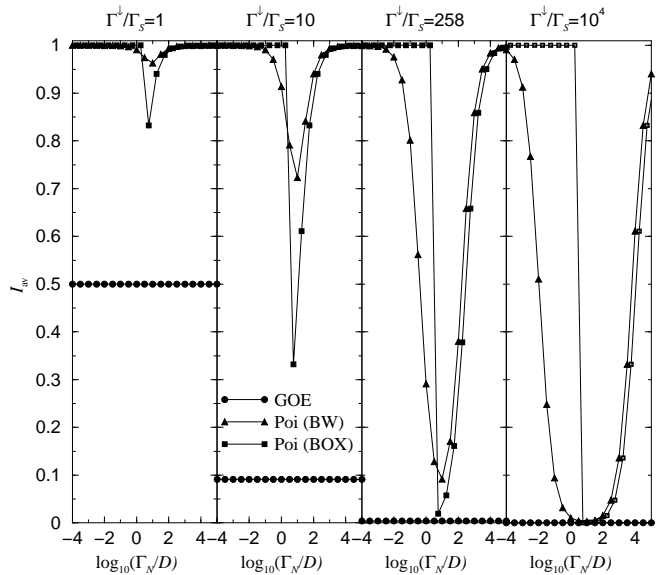}
\caption{\label{fig3} Decay intensity $I_\av$ vs. $\log_{10}(\Ga_N/D)$
for $\De/D$=0.25. See also caption to Fig.~\ref{fig1}.}
\end{figure}

Eq.~(\ref{Iav}) for $I_\av$  depends on 
four parameters: $\Ga_S$, $|V_{0d}|^2$, $\Ga_N$ and
the distance in energy separating $|d\ket$ from $|0\ket$, 
$\De=E_d-E_0$. It is useful to introduce a spreading width 
defined by $\Ga^\da=2\pi|V_{0d}|^2/D$.
Upon making the change of integration variable 
$x=(E-E_0)/D$, Eq.~(\ref{Iav}) takes the form (we set the energy shift
$R_d(E)=0$ as doing so does not modify our conclusions)
\bea\nonumber
&&I_\av=\f{\Ga_S/D}{2\pi}\int_{-\infty}^{\infty}dx
\\&&\f{1}{x^2+\left(\Ga_S/D\right)^2\left[1
+\f{\Ga^\da}{\Ga_S}S_d(Dx+E_0)\right]^2/4}.
\label{1Iav}
\eea
Inserting Eq.~(\ref{BOX}) for $S_d$ into Eq.~(\ref{1Iav}) we obtain
\bea\nonumber
&&I_\av^{\rm{BOX}}=1+\f{1}{\pi}\biggl[\arctan \theta_- -\arctan\theta_+
\biggr.
\\&+&\biggl.\f{1}{1+\f{2\Ga^\da/\Ga_S}{\pi\Ga_N/D}}
\left\{\arctan \phi_+ -\arctan\phi_-\right\}
\biggr],
\label{IavBOX}
\eea 
where
\be\label{chi1}
\theta_\pm=\f{1}{\Ga_S/D}\left[2\De/D\pm\pi\Ga_N/D\right]
\ee
and
\be\label{chi2}
\phi_\pm=\f{\theta_\pm}{1+\f{2\Ga^\da/\Ga_S}{\pi\Ga_N/D}}.
\ee

\begin{figure}
\includegraphics[width=.48\textwidth]{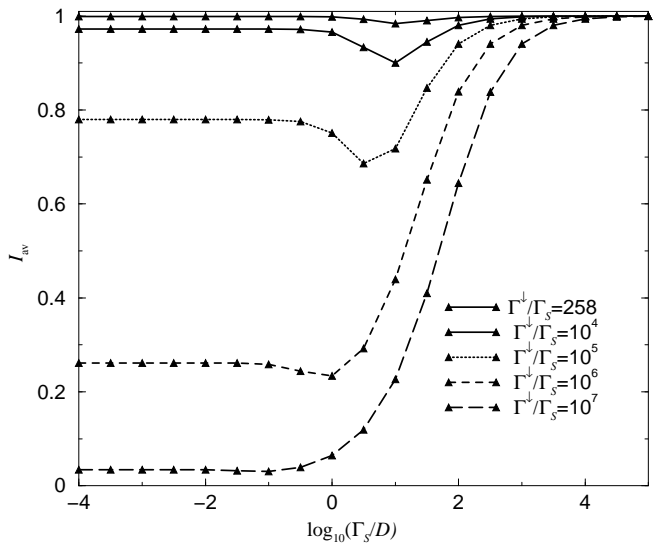}
\caption{\label{fig4} Decay intensity $I_\av$ vs. $\log_{10}(\Ga_S/D)$.
See also caption to Fig.~\ref{fig1}.}
\end{figure}
From Eqs.~(\ref{IavBOX}-\ref{chi2}) it is seen that as $\la\rightarrow 0$, 
$I_\av$ is a function of four dimensionless variables:
$\Ga^\da/\Ga_S$, $\Ga_N/D$, $\Ga_S/D$ and $\De/D$. 
Figures~\ref{fig1}, \ref{fig2}, \ref{fig3} and \ref{fig4} show $I_\av$ vs. 
$\De/D$, $\Ga^\da/\Ga_S$ and $\Ga_N/D$ and $\Ga_S/D$ respectively.
For the Poisson limit a 
 significant dependence of $I_\av$ on all four parameters is observed.
In all the graphs $D$=16.3 eV, $\Ga_N$=$4.8\times 10^{-3}$ eV and 
$\Ga_S$=$97\times 10^{-6}$ eV which are the values for $^{194}$Hg-1 at
spin 12 $\hbar$ \cite{Krucken:2001we}. 
The triangles and squares represent the Breit-Wigner and box function 
representations of the  Poisson limit respectively,
whilst the circles represent the GOE limit. The decay-out is enhanced
by increasing the degree of chaos if the triangles or squares are
above the circles and it is hindered
if the triangles or squares are below the circles. 

The authors of Ref.~\cite{Krucken:2001we} obtained
a spreading width of $\Ga^\da$=0.025 eV ($\Ga^\da/\Ga_S$=258)
from an experimental value for the total intraband decay intensity 
at spin 12 $\hbar$ equal 
to 0.58, using the theory of Ref.~\cite{Gu:1999bv}. They assume
that the fluctuation contribution \cite{endnote15} dominates 
($I_\av^{\rm{GOE}}$=1/259). 
It is clear from Fig.~\ref{fig1} that the extraction of $\Ga^\da$ from
experimental data using the results for the Poisson limit of the 
present paper would be 
extremely sensitive to $\De$. The energy difference $\De$ is an additional
unknown parameter. 

From Figs.~\ref{fig1} and \ref{fig2} we see that increasing the 
chaoticity, $\la$, from 0 to 1 only hinders the decay-out when $\De\sim 0$
as was observed by \AA berg \cite{Aberg:2003}.
However, even though a chaos enhancement is obtained for $\De$ 
sufficiently large,
it is more convincing to explain the decay-out by an increase in 
$\Ga^\da/\Ga_S$, than by an increase in $\la$, for the following reasons: 
Firstly, we see from Figs.~\ref{fig1} to \ref{fig3}
that increasing $\la$ from 0 to 1 cannot exhaust all
of the intra-band decay intensity unless 
$\Ga^\da/\Ga_S\rightarrow\infty$. Indeed, the extent to which
the an order-chaos transition may exhaust the intra-band decay intensity 
is determined solely by $\Ga^\da/\Ga_S$ 
[$I_\av^{\rm{GOE}}=(1+\Ga^\da/\Ga_S)^{-1}$]. 
For example, when 
$\Ga^\da/\Ga_S=0.01$, an order-chaos transition will reduce (if $\De\gg 0$) 
$I_\av$ from 1 to 0.99 - a rather small effect. 
Further, when $\Ga^\da/\Ga_S=0$ it is impossible 
for an increase of chaos to trigger the decay-out since $I_\av=1$ for 
values of $\la$ [see Eq.~(\ref{1Iav})];
Secondly, a chaos-order transition is not necessary to trigger the 
decay-out since $I_\av\rightarrow 0$ as $\Ga^\da/\Ga_S\rightarrow\infty$
whatever the values of $\la$ and $\De$ as long as $\Ga_N$ has a non-zero 
value [see Eq.~(\ref{1Iav})]. 
It may be seen from Fig.~\ref{fig2} that this is the case 
even for $\la=0$ (when $S_d(E)$ is described by the Breit-Wigner function).

It is true to say that $\la$ modifies the spin at which the
decay-out occurs as can be seen from Fig.~\ref{fig2} (see also Fig.~3
of Ref.~\cite{Sargeant:2001aq}). However the arguments of the preceding
paragraph convince us that the spin dependence of $\Ga^\da/\Ga_S$ is of
more importance.
Since $\Ga^\da/\Ga_S$ is determined by the deformation barrier these
arguments reinforce our belief in the conclusion of 
Ref.~\cite{Sargeant:2001aq} 
that the attenuation of the intra-band intensity with decreasing spin
is mostly due to the spin dependence of the barrier.

\bibliographystyle{apsrev}
\bibliography{sd,sargeant,rmt}

\begin{thebibliography}{14}
\expandafter\ifx\csname natexlab\endcsname\relax\def\natexlab#1{#1}\fi
\expandafter\ifx\csname bibnamefont\endcsname\relax
  \def\bibnamefont#1{#1}\fi
\expandafter\ifx\csname bibfnamefont\endcsname\relax
  \def\bibfnamefont#1{#1}\fi
\expandafter\ifx\csname citenamefont\endcsname\relax
  \def\citenamefont#1{#1}\fi
\expandafter\ifx\csname url\endcsname\relax
  \def\url#1{\texttt{#1}}\fi
\expandafter\ifx\csname urlprefix\endcsname\relax\def\urlprefix{URL }\fi
\providecommand{\bibinfo}[2]{#2}
\providecommand{\eprint}[2][]{\url{#2}}

\bibitem[{\citenamefont{{Wilson} et~al.}(2003)\citenamefont{{Wilson},
  {Dracoulis}, {Byrne}, {Davidson}, {Lane}, {Clark}, {Fallon}, {G{\" o}rgen},
  {Macchiavelli}, and {Ward}}}]{Wilson:2003}
\bibinfo{author}{\bibfnamefont{A.~N.} \bibnamefont{{Wilson}}},
  \bibinfo{author}{\bibfnamefont{G.~D.} \bibnamefont{{Dracoulis}}},
  \bibinfo{author}{\bibfnamefont{A.~P.} \bibnamefont{{Byrne}}},
  \bibinfo{author}{\bibfnamefont{P.~M.} \bibnamefont{{Davidson}}},
  \bibinfo{author}{\bibfnamefont{G.~J.} \bibnamefont{{Lane}}},
  \bibinfo{author}{\bibfnamefont{R.~M.} \bibnamefont{{Clark}}},
  \bibinfo{author}{\bibfnamefont{P.}~\bibnamefont{{Fallon}}},
  \bibinfo{author}{\bibfnamefont{A.}~\bibnamefont{{G{\" o}rgen}}},
  \bibinfo{author}{\bibfnamefont{A.~O.} \bibnamefont{{Macchiavelli}}},
  \bibnamefont{and} \bibinfo{author}{\bibfnamefont{D.}~\bibnamefont{{Ward}}},
  \bibinfo{journal}{Phys. Rev. Lett.} \textbf{\bibinfo{volume}{90}},
  \bibinfo{pages}{142501} (\bibinfo{year}{2003}).

\bibitem[{\citenamefont{{Lauritsen} et~al.}(2002)\citenamefont{{Lauritsen},
  {Carpenter}, {D{\o}ssing}, {Fallon}, {Herskind}, {Janssens}, {Jenkins},
  {Khoo}, {Kondev}, {Lopez-Martens} et~al.}}]{Lauritsen:2002}
\bibinfo{author}{\bibfnamefont{T.}~\bibnamefont{{Lauritsen}}},
  \bibinfo{author}{\bibfnamefont{M.~P.} \bibnamefont{{Carpenter}}},
  \bibinfo{author}{\bibfnamefont{T.}~\bibnamefont{{D{\o}ssing}}},
  \bibinfo{author}{\bibfnamefont{P.}~\bibnamefont{{Fallon}}},
  \bibinfo{author}{\bibfnamefont{B.}~\bibnamefont{{Herskind}}},
  \bibinfo{author}{\bibfnamefont{R.~V.} \bibnamefont{{Janssens}}},
  \bibinfo{author}{\bibfnamefont{D.~G.} \bibnamefont{{Jenkins}}},
  \bibinfo{author}{\bibfnamefont{T.~L.} \bibnamefont{{Khoo}}},
  \bibinfo{author}{\bibfnamefont{F.~G.} \bibnamefont{{Kondev}}},
  \bibinfo{author}{\bibfnamefont{A.}~\bibnamefont{{Lopez-Martens}}},
  \bibnamefont{et~al.}, \bibinfo{journal}{Phys. Rev. Lett.}
  \textbf{\bibinfo{volume}{88}}, \bibinfo{pages}{042501}
  (\bibinfo{year}{2002}).

\bibitem[{\citenamefont{{Paul} et~al.}(2001)\citenamefont{{Paul}, {Forbes},
  {Gizon}, {Hauschild}, {Hibbert}, {Joss}, {Nolan}, {Nyak{\' o}}, {Sampson},
  {Semple} et~al.}}]{Paul:2001}
\bibinfo{author}{\bibfnamefont{E.~S.} \bibnamefont{{Paul}}},
  \bibinfo{author}{\bibfnamefont{S.~A.} \bibnamefont{{Forbes}}},
  \bibinfo{author}{\bibfnamefont{J.}~\bibnamefont{{Gizon}}},
  \bibinfo{author}{\bibfnamefont{K.}~\bibnamefont{{Hauschild}}},
  \bibinfo{author}{\bibfnamefont{I.~M.} \bibnamefont{{Hibbert}}},
  \bibinfo{author}{\bibfnamefont{D.~T.} \bibnamefont{{Joss}}},
  \bibinfo{author}{\bibfnamefont{P.~J.} \bibnamefont{{Nolan}}},
  \bibinfo{author}{\bibfnamefont{B.~M.} \bibnamefont{{Nyak{\' o}}}},
  \bibinfo{author}{\bibfnamefont{J.~A.} \bibnamefont{{Sampson}}},
  \bibinfo{author}{\bibfnamefont{A.~T.} \bibnamefont{{Semple}}},
  \bibnamefont{et~al.}, \bibinfo{journal}{Nucl. Phys.}
  \textbf{\bibinfo{volume}{A690}}, \bibinfo{pages}{341} (\bibinfo{year}{2001}).

\bibitem[{\citenamefont{{Dewald} et~al.}(2001)\citenamefont{{Dewald}, {K{\"
  u}hn}, {Peusquens}, {von Brentano}, {Kr{\" u}cken}, {Deleplanque}, {Lee},
  {Clark}, {Fallon}, {Macchiavelli} et~al.}}]{Dewald:2001}
\bibinfo{author}{\bibfnamefont{A.}~\bibnamefont{{Dewald}}},
  \bibinfo{author}{\bibfnamefont{R.}~\bibnamefont{{K{\" u}hn}}},
  \bibinfo{author}{\bibfnamefont{R.}~\bibnamefont{{Peusquens}}},
  \bibinfo{author}{\bibfnamefont{P.}~\bibnamefont{{von Brentano}}},
  \bibinfo{author}{\bibfnamefont{R.}~\bibnamefont{{Kr{\" u}cken}}},
  \bibinfo{author}{\bibfnamefont{M.~A.} \bibnamefont{{Deleplanque}}},
  \bibinfo{author}{\bibfnamefont{I.~Y.} \bibnamefont{{Lee}}},
  \bibinfo{author}{\bibfnamefont{R.~M.} \bibnamefont{{Clark}}},
  \bibinfo{author}{\bibfnamefont{P.}~\bibnamefont{{Fallon}}},
  \bibinfo{author}{\bibfnamefont{A.~O.} \bibnamefont{{Macchiavelli}}},
  \bibnamefont{et~al.}, \bibinfo{journal}{Phys. Rev. C}
  \textbf{\bibinfo{volume}{64}}, \bibinfo{pages}{054309}
  (\bibinfo{year}{2001}).

\bibitem[{\citenamefont{{Yoshida} et~al.}(2001)\citenamefont{{Yoshida},
  {Matsuo}, and {Shimizu}}}]{Yoshida:2001}
\bibinfo{author}{\bibfnamefont{K.}~\bibnamefont{{Yoshida}}},
  \bibinfo{author}{\bibfnamefont{M.}~\bibnamefont{{Matsuo}}}, \bibnamefont{and}
  \bibinfo{author}{\bibfnamefont{Y.~R.} \bibnamefont{{Shimizu}}},
  \bibinfo{journal}{Nucl. Phys.} \textbf{\bibinfo{volume}{A696}},
  \bibinfo{pages}{85} (\bibinfo{year}{2001}).

\bibitem[{\citenamefont{{{\AA}berg}}(1999)}]{Aberg:1999a}
\bibinfo{author}{\bibfnamefont{S.}~\bibnamefont{{{\AA}berg}}},
  \bibinfo{journal}{Phys. Rev. Lett.} \textbf{\bibinfo{volume}{82}},
  \bibinfo{pages}{299} (\bibinfo{year}{1999}).

\bibitem[{\citenamefont{Sargeant
  et~al.}(2002{\natexlab{a}})\citenamefont{Sargeant, Hussein, Pato, Takigawa,
  and Ueda}}]{Sargeant:2001aq}
\bibinfo{author}{\bibfnamefont{A.~J.} \bibnamefont{Sargeant}},
  \bibinfo{author}{\bibfnamefont{M.~S.} \bibnamefont{Hussein}},
  \bibinfo{author}{\bibfnamefont{M.~P.} \bibnamefont{Pato}},
  \bibinfo{author}{\bibfnamefont{N.}~\bibnamefont{Takigawa}}, \bibnamefont{and}
  \bibinfo{author}{\bibfnamefont{M.}~\bibnamefont{Ueda}},
  \bibinfo{journal}{Phys. Rev. C} \textbf{\bibinfo{volume}{65}},
  \bibinfo{pages}{024302} (\bibinfo{year}{2002}{\natexlab{a}}).

\bibitem[{\citenamefont{{{\AA}berg}}(2003)}]{Aberg:2003}
\bibinfo{author}{\bibfnamefont{S.}~\bibnamefont{{{\AA}berg}}},
  \bibinfo{journal}{Phys. Rev. C} \textbf{\bibinfo{volume}{68}},
  \bibinfo{pages}{069801} (\bibinfo{year}{2003}).

\bibitem[{\citenamefont{Krucken et~al.}(2001)\citenamefont{Krucken, Dewald, von
  Brentano, and Weidenmuller}}]{Krucken:2001we}
\bibinfo{author}{\bibfnamefont{R.}~\bibnamefont{Krucken}},
  \bibinfo{author}{\bibfnamefont{A.}~\bibnamefont{Dewald}},
  \bibinfo{author}{\bibfnamefont{P.}~\bibnamefont{von Brentano}},
  \bibnamefont{and} \bibinfo{author}{\bibfnamefont{H.~A.}
  \bibnamefont{Weidenmuller}}, \bibinfo{journal}{Phys. Rev. C}
  \textbf{\bibinfo{volume}{64}}, \bibinfo{pages}{064316}
  (\bibinfo{year}{2001}).

\bibitem[{\citenamefont{Sargeant et~al.}(2000)\citenamefont{Sargeant, Hussein,
  Pato, and Ueda}}]{Sargeant:1999qk}
\bibinfo{author}{\bibfnamefont{A.~J.} \bibnamefont{Sargeant}},
  \bibinfo{author}{\bibfnamefont{M.~S.} \bibnamefont{Hussein}},
  \bibinfo{author}{\bibfnamefont{M.~P.} \bibnamefont{Pato}}, \bibnamefont{and}
  \bibinfo{author}{\bibfnamefont{M.}~\bibnamefont{Ueda}},
  \bibinfo{journal}{Phys. Rev. C} \textbf{\bibinfo{volume}{61}},
  \bibinfo{pages}{011302} (\bibinfo{year}{2000}).

\bibitem[{\citenamefont{Weidenmuller et~al.}(1998)\citenamefont{Weidenmuller,
  von Brentano, and Barrett}}]{Weidenmuller:1998xf}
\bibinfo{author}{\bibfnamefont{H.~A.} \bibnamefont{Weidenmuller}},
  \bibinfo{author}{\bibfnamefont{P.}~\bibnamefont{von Brentano}},
  \bibnamefont{and} \bibinfo{author}{\bibfnamefont{B.~R.}
  \bibnamefont{Barrett}}, \bibinfo{journal}{Phys. Rev. Lett.}
  \textbf{\bibinfo{volume}{81}}, \bibinfo{pages}{3603} (\bibinfo{year}{1998}).

\bibitem[{\citenamefont{Gu and Weidenm{\"u}ller}(1999)}]{Gu:1999bv}
\bibinfo{author}{\bibfnamefont{J.-z.} \bibnamefont{Gu}} \bibnamefont{and}
  \bibinfo{author}{\bibfnamefont{H.~A.} \bibnamefont{Weidenm{\"u}ller}},
  \bibinfo{journal}{Nucl. Phys.} \textbf{\bibinfo{volume}{A660}},
  \bibinfo{pages}{197} (\bibinfo{year}{1999}).

\bibitem[{\citenamefont{Sargeant
  et~al.}(2002{\natexlab{b}})\citenamefont{Sargeant, Hussein, Pato, and
  Ueda}}]{Sargeant:2002sv}
\bibinfo{author}{\bibfnamefont{A.~J.} \bibnamefont{Sargeant}},
  \bibinfo{author}{\bibfnamefont{M.~S.} \bibnamefont{Hussein}},
  \bibinfo{author}{\bibfnamefont{M.~P.} \bibnamefont{Pato}}, \bibnamefont{and}
  \bibinfo{author}{\bibfnamefont{M.}~\bibnamefont{Ueda}},
  \bibinfo{journal}{Phys. Rev. C} \textbf{\bibinfo{volume}{66}},
  \bibinfo{pages}{064301} (\bibinfo{year}{2002}{\natexlab{b}}).

\bibitem[{\citenamefont{Hussein et~al.}(2003)\citenamefont{Hussein, Sargeant,
  Pato, Takigawa, and Ueda}}]{Hussein:2003ia}
\bibinfo{author}{\bibfnamefont{M.~S.} \bibnamefont{Hussein}},
  \bibinfo{author}{\bibfnamefont{A.~J.} \bibnamefont{Sargeant}},
  \bibinfo{author}{\bibfnamefont{M.~P.} \bibnamefont{Pato}},
  \bibinfo{author}{\bibfnamefont{N.}~\bibnamefont{Takigawa}}, \bibnamefont{and}
  \bibinfo{author}{\bibfnamefont{M.}~\bibnamefont{Ueda}}
  (\bibinfo{year}{2003}), \bibinfo{note}{to appear in Progress of Theoretical
  Physics}, \eprint{nucl-th/0312129}.

\end{thebibliography}

\end{document}